\documentclass[12pt]{article}
\usepackage{graphicx}
\begin{document}

\begin{center}

{\Large \bf  Lorentz Harmonics, Squeeze Harmonics, and \\[0.5ex]
their Physical Applications}\\
\vspace{2ex}

Young S. Kim~\footnote{email: yskim@umd.edu}\\
Center for Fundamental Physics, University of Maryland, \\
College Park, Maryland 20742, U.S.A.\\
\vspace{2ex}
Marilyn E. Noz~\footnote{email: marilyne.noz@gmail.com} \\
Department of Radiology, New York University,\\
New York, New York 10016, U.S.A. \\

\end{center}

\vspace{4ex}

\begin{abstract}
Among the symmetries in physics, the rotation symmetry is most
familiar to us.  It is known that the spherical harmonics
serve useful purposes when the world is rotated.  Squeeze
transformations are also becoming more prominent in physics,
particularly in optical sciences and in high-energy physics.
As can be seen from Dirac's light-cone coordinate system,
Lorentz boosts are squeeze transformations.  Thus the squeeze
transformation is one of the fundamental transformations in
Einstein's Lorentz-covariant world.  It is possible to define
a complete set of orthonormal functions defined for one Lorentz
frame.  It is shown that the same set can be used for other
Lorentz frames.  Transformation properties are discussed.
Physical applications are discussed in both optics and
high-energy physics.  It is shown that the Lorentz harmonics
provide the mathematical basis for squeezed states of light.
It is shown also that the same set of harmonics can be used for
understanding Lorentz-boosted hadrons in high-energy physics.
It is thus possible to transmit physics from one branch of
physics to the other branch using the mathematical basis
common to them.

\end{abstract}

\newpage

\section{Introduction}\label{intro}

In this paper, we are concerned with symmetry transformations
in two dimensions, and we are accustomed to the coordinate system
specified by $x$ and $y$ variables.  On the $xy$ plane, we know
how to make rotations and translations.   The rotation in the
$xy$ plane is performed by the matrix algebra
\begin{equation}\label{rot01}
  \pmatrix{x'\cr y'} = \pmatrix{\cos\theta  &  -\sin\theta \cr
               \sin\theta  & \cos\theta}  \pmatrix{x \cr y } ,
\end{equation}
but we are not yet familiar with
\begin{equation}\label{sqz01}
  \pmatrix{z'\cr t'} = \pmatrix{\cosh\eta  &  \sinh\eta \cr
               \sinh\eta  & \cosh\eta}  \pmatrix{z \cr t } .
\end{equation}
We see this form when we learn Lorentz transformations, but there
is a tendency in the literature to avoid this form, especially in
high-energy physics.  Since this transformation can also be written
as
\begin{equation}\label{sqz02}
  \pmatrix{u' \cr v'} = \pmatrix{\exp{(\eta)}  &  0 \cr
              0  & \exp{(-\eta)}} \pmatrix{u \cr v } ,
\end{equation}
with
\begin{equation}\label{licone01}
 u = \frac{z + t}{\sqrt{2}}, \qquad  v = \frac{z - t}{\sqrt{2}} ,
\end{equation}
where the variables $u$ and $v$ are expanded and contracted
respectively, we call Eq.(\ref{sqz01}) or Eq.(\ref{sqz02}) {\bf squeeze}
transformations~\cite{knp91}.

\par

From the mathematical point of view, the symplectic group $Sp(2)$
contains both the rotation and squeeze transformations of
Eqs.~(\ref{rot01}) and ~(\ref{sqz01}), and its mathematical properties
have been extensively discussed in the literature~\cite{knp91,guil84}.
This group has been shown to be one of the essential tools in quantum
optics.  From the mathematical point of view, the squeezed state in
quantum optics is a harmonic oscillator representation of this $Sp(2)$
group~\cite{knp91}.

\par

We are interested in this paper in ``squeeze transformations'' of
localized functions.  We are quite familiar with the role of
spherical harmonics in three dimensional rotations.  We use there
the same set of harmonics, but the rotated function has different
linear combinations of those harmonics.  Likewise, we are
interested in a complete set of functions which will serve
the same purpose for squeeze transformations.  It will be shown
that harmonic oscillator wave functions can serve the desired
purpose.  From the physical point of view, squeezed states define
the squeeze or Lorentz harmonics.

\par
In 2003, Giedke {\it et al.} used the Gaussian function to discuss
the entanglement problems in information theory~\cite{gied03}.
This paper allows us to use the oscillator wave functions to
address many interesting current issues in quantum optics and
information theory.  In 2005, the present authors noted that the
formalism of Lorentz-covariant harmonic oscillators leads to a
space-time entanglement~\cite{kn05job}.  We developed the oscillator
formalism to deal with hadronic phenomena observed in high-energy
laboratories~\cite{knp86}.  It is remarkable that the mathematical
formalism of Giedke {\it et al.} is identical with that of our
oscillator formalism.

\par

While quantum optics or information theory is a relatively new
branch of physics, the squeeze transformation has been the backbone
of Einstein's special relativity.  While Lorentz, Poincar\'e, and
Einstein used the transformation of Eq.(\ref{sqz01}) for Lorentz
boosts, Dirac observed that the same equation can be written in the
form of Eq.(\ref{sqz02})~\cite{dir49}.  Unfortunately, this squeeze
aspect of Lorentz boosts has not been fully addressed in high-energy
physics dealing with particles moving with relativistic speeds.

\par
Thus, we can call the same set of functions ``squeeze harmonics''
and ``Lorentz harmonics'' in quantum optics and high-energy
physics respectively.  This allows us to translate the physics
of quantum optics or information theory into that of high-energy
physics.

\par

The physics of high-energy hadrons requires a Lorentz-covariant
localized quantum system.  This description requires one
variable which is hidden in the present form of quantum mechanics.
It is the time-separation variable between two constituent
particles in a quantum bound system like the hydrogen atom, where
the Bohr radius measures the separation between the proton and
the electron.  What happens to this quantity when the hydrogen
atom is boosted and the time-separation variable starts playing
its role?  The Lorentz harmonics will allow us to address this
question.

\par

In Sec.~\ref{harmonics}, it is noted that the Lorentz boost of
localized wave functions can be described in terms of one-dimensional
harmonic oscillators.  Thus, those wave functions constitute the
Lorentz harmonics.  It is also noted that the Lorentz boost is a
squeeze transformation.
\par

In Sec.~\ref{diracqm}, we examine Dirac's life-long efforts to make
quantum mechanics consistent with special relativity, and present a
Lorentz-covariant form of bound-state quantum mechanics.
In Sec.~\ref{proba}, we construct a set of Lorentz-covariant
harmonic oscillator wave functions, and show that they can be given
a Lorentz-covariant probability interpretation.
\par

In Sec.~\ref{squeeze}, the formalism is shown to constitute
a mathematical basis for squeezed states of light, and for quantum
entangled states.  In Sec.~\ref{restof}, this formalism can serve
as the language for Feynman's rest of the universe~\cite{fey72}.
Finally, in Sec.~\ref{quarkmo}, we show that the harmonic oscillator
formalism can be applied to high-energy hadronic physics, and what
we observe there can be interpreted in terms of what we learn from
quantum optics.

\section{Lorentz or Squeeze Harmonics}\label{harmonics}

Let us start with the two-dimensional plane.  We are quite familiar
with rigid transformations such as rotations and translations in
two-dimensional space.  Things are different for non-rigid
transformations such as a circle becoming an ellipse.
\par

We start with the well-known one-dimensional harmonic oscillator
eigenvalue equation
\begin{equation}\label{osc01}
\frac{1}{2} \left[-\left(\frac{\partial} {\partial x} \right)^2
  + x^2 \right] \chi_n (x) = \left(n + \frac{1}{2}\right)\chi_n(x) .
\end{equation}
For a given value of integer $n,$ the solution takes the form
\begin{equation}\label{owf01}
     \chi_n (x) = \left[\frac{1}{\sqrt{\pi}2^n n!}\right]^{1/2}
              H_n(x) \exp{\left(\frac{-x^2}{2}\right)} ,
\end{equation}
where $H_n(x)$ is the Hermite polynomial of the n-th degree.  We can
then consider a set of functions with all integer values of $n$.  They
satisfy the orthogonality relation
\begin{equation}
  \int \chi_n(x) \chi_{n'}(x) = \delta_{nn'} .
\end{equation}
This relation allows us to define $f(x)$ as
\begin{equation}
  f(x) = \sum_{n} A_n \chi_{n}(x),
\end{equation}
with
\begin{equation}
  A_n = \int f(x) \chi_{n} (x) dx .
\end{equation}

\par

Let us next consider another variable added to Eq.(\ref{osc01}), and
the differential equation
\begin{equation}\label{osc02}
\frac{1}{2} \left\{\left[-\left(\frac{\partial}
 {\partial x} \right)^2 + x^2 \right]
 + \left[-\left(\frac{\partial}{\partial y}\right)^2 +
 y^2\right]\right\} \phi(x,y) = \lambda \phi(x,y) ,
\end{equation}
This equation can be re-arranged to
\begin{equation}\label{osc03}
\frac{1}{2} \left\{-\left(\frac{\partial} {\partial x}\right)^2
 -\left(\frac{\partial}{\partial y}\right)^2 + x^2  +  y^2\right\}
 \phi(x,y) = \lambda \phi(x,y) ,
\end{equation}

\par

This differential equation is invariant under the rotation defined in
Eq.(\ref{rot01}).  In terms of the polar coordinate system with
\begin{equation}
r = \sqrt{x^2 + y^2}, \qquad \tan\theta = \left(\frac{y}{x}\right)
\end{equation}
this equation can be written:
\begin{equation}\label{osc10}
\frac{1}{2} \left\{ -\frac{\partial^2}{\partial r ^2 }
 -\frac{1}{r}\frac{\partial}{\partial r} -
    \frac{1}{r^2} \frac{\partial^2}{\partial \theta^2}
 + r^2 \right\} \phi(r,\theta) = \lambda \phi(r,\theta) ,
\end{equation}
and the solution takes the form
\begin{equation}\label{polar01}
\phi(r,\theta) = e^{-r^2/2} R_{n,m}(r)\left\{A_m\cos(m\theta) +
                              B_n\sin(m\theta)\right\} .
\end{equation}
The radial equation should satisfy
\begin{equation}\label{osc12}
\frac{1}{2} \left\{ -\frac{\partial^2}{\partial r ^2 }
 -\frac{1}{r}\frac{\partial}{\partial r} + \frac{m^2}{r^2}
      + r^2 \right\} R_{n,m}(r) = (n + m + 1) R_{n,m}(r) .
\end{equation}
In the polar form of Eq.(\ref{polar01}), we can achieve the
rotation of this function by changing the angle variable $\theta$.

\par

On the other hand, the differential equation of Eq.(\ref{osc02})
is separable in the $x$ and $y$ variables.  The eigen solution
takes the form
\begin{equation}\label{sol01}
\phi_{n_x,n_y}(x,y) = \chi_{n_x}(x) \chi_{n_y}(y),
\end{equation}
with
\begin{equation}
\lambda = n_x + n_y + 1 .
\end{equation}
If a function
$f(x,y)$ is sufficiently localized around the origin, it can be
expanded as
\begin{equation}
  f(x,y) = \sum_{n_x,n_y} A_{n_x,n_y} \chi_{n_x}(x) \chi_{n_y}(y),
\end{equation}
with
\begin{equation}
  A_{n_x,n_y} = \int f(x,y)\chi_{n_x}(x)  \chi_{n_y}(y)~dx~dy.
\end{equation}
If we rotate $f(x,y)$ according to Eq.(\ref{rot01}), it becomes
$f(x^*, y^*)$, with
\begin{equation}
x^* = (\cos\theta) x - (\sin\theta) y, \qquad
    y^* = (\sin\theta) x + (\cos\theta) y
\end{equation}
This rotated function can also be expanded in terms of
$\chi_{n_x}(x)$ and $\chi_{n_y}(y)$:
\begin{equation}
  f(x^*,y^*) = \sum_{n_x,n_y} A^*_{n_x,n_y} \chi_{n_x}(x) \chi_{n_y}(y),
\end{equation}
with
\begin{equation}
  A^*_{n_x,n_y} = \int f(x^*,y^*)\chi_{n_x}(x)  \chi_{n_y}(y)~dx~dy.
\end{equation}

   \par
Next, let us consider the differential equation
\begin{equation}\label{osc05}
\frac{1}{2} \left\{-\left(\frac{\partial} {\partial z}\right)^2
 + \left(\frac{\partial}{\partial t}\right)^2   +
  z^2  -  t^2\right\} \psi (z,t) = \lambda \psi (z,t) .
\end{equation}
Here we use the variables $z$ and $t$, instead of $x$ and $y$.
Clearly, this equation can be also separated in the $z$ and $t$
coordinates, and the eigen solution can be written as
\begin{equation}\label{sol02}
\psi_{n_z,n_t}(z,t) = \chi_{n_z}(z) \chi_{n_t}(z,t),
\end{equation}
with
\begin{equation}
\lambda = n_z - n_t.
\end{equation}
\par
The oscillator equation is not invariant under coordinate rotations of
the type given in Eq.(\ref{rot01}).  It is however invariant under the
squeeze transformation given in Eq.(\ref{sqz01}).

\par

The differential equation of Eq.(\ref{osc05}) becomes
\begin{equation}\label{sqz05}
\frac{1}{4} \left\{-\frac{\partial} {\partial u}\frac{\partial} {\partial v}
 + uv \right\} \psi (u,v) = \lambda \psi (u,v) .
\end{equation}
\par
Both Eq.(\ref{osc03}) and Eq.(\ref{osc05}) are two-dimensional
differential equations. They are invariant under rotations and squeeze
transformations respectively.  They take convenient forms in the polar and
squeeze coordinate systems respectively as shown in Eq.(\ref{osc10}) and
Eq.(\ref{sqz05}).
\par
The solutions of the rotation-invariant equation are well known, but the
solutions of the squeeze-invariant equation are still strange to the
physics community.  Fortunately, both equations are separable in the Cartesian
coordinate  system.  This allows us to study the latter in terms of the
familiar rotation-invariant equation.  This means that if the solution is
sufficiently localized in the $z$ and $t$ plane, it can be written as
\begin{equation}
  \psi(z,t) = \sum_{n_z,n_t} A_{n_z,n_t} \chi_{n_z}(z) \chi_{n_t}(t),
\end{equation}
with
\begin{equation}
  A_{n_z,n_t} = \int \psi(z,t)\chi_{n_z}(z) \chi_{n_t}(t)~dz~dt.
\end{equation}
If we squeeze the coordinate according to Eq.(\ref{sqz01}),
\begin{equation}
  \psi(z^*,t^*) = \sum_{n_z,n_t} A^*_{n_z,n_t} \chi_{n_z}(z) \chi_{n_t}(t),
\end{equation}
with
\begin{equation}
  A^*_{n_z,n_t} = \int \psi(z^*,t^*)\chi_{n_z}(z) \chi_{n_t}(t)~dz~dt.
\end{equation}
Here again both the original and transformed wave functions are linear
combinations of the wave functions for the one-dimensional harmonic
oscillator given in Eq.(\ref{owf01}).

\par
The wave functions for the one-dimensional oscillator are well known, and
they play important roles in many branches of physics.  It is gratifying
to note that they could play an essential role in squeeze transformations
and Lorentz boosts.  We choose to call them Lorentz harmonics or
squeeze harmonics.

\begin{table}[th]
\caption{Cylindrical and hyperbolic equations.  The cylindrical equation
is invariant under rotation while the hyperbolic equation is invariant
under squeeze transformation}\label{kimnoz}
\vspace{3mm}
\begin{center}
\begin{tabular}{clccc}
\hline
\hline
{} &{} &  {} & {}& {}\\
{} &  Equation  & Invariant under & Eigenvalue & {}
\\[4mm]\hline
{} & {} & {} & {}\\
{} & Cylindrical & Rotation & $\lambda = n_x + n_y + 1 $ & {}
\\[4mm]\hline
{} &  {} & {} & {}\\
{} &   Hyperbolic & Squeeze   & $\lambda = n_x - n_y $ & {}
\\[4mm]
\hline
\hline
\end{tabular}
\end{center}
\end{table}

\section{The Physical Origin of Squeeze Transformations}\label{diracqm}

Paul A. M. Dirac made it his life-long effort to combine quantum mechanics
with special relativity.  We examine the following four of his papers.

\begin{itemize}

\item In 1927~\cite{dir27}, Dirac pointed out the time-energy uncertainty
   should be taken into consideration for efforts to combine quantum
   mechanics and special relativity.

\item In 1945~\cite{dir45}, Dirac considered four-dimensional harmonic
    oscillator wave functions with
    \begin{equation}\label{owf02}
  \exp{\left\{ - \frac{1}{2}\left(x^2 + y^2 + z^2 + t^2\right)\right\}} ,
    \end{equation}
      and noted that this form is not Lorentz-covariant.

\item In 1949~\cite{dir49}, Dirac introduced the light-cone variables of
    Eq.(\ref{licone01}).  He also noted that the construction of a
    Lorentz-covariant quantum mechanics is equivalent to the construction
    of a representation of the Poncar\'e group.

\item In 1963~\cite{dir63}, Dirac constructed a representation of the
   (3 + 2) deSitter group using two harmonic oscillators.  This deSitter
   group contains three (3 + 1) Lorentz groups as its subgroups.
\end{itemize}

\begin{figure}
\centerline{\includegraphics[scale=0.43]{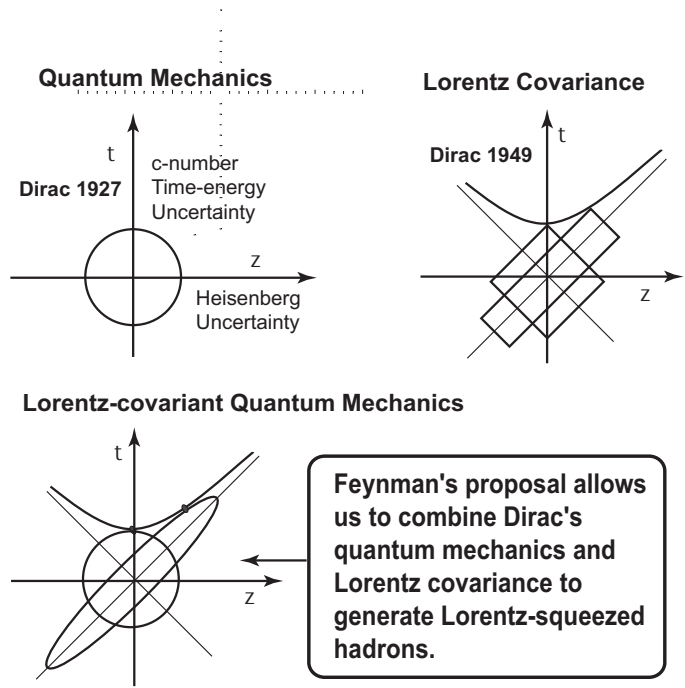}}
\caption{Space-time picture of quantum mechanics.  In his 1927 paper,
Dirac noted that there is a c-number time-energy uncertainty
relation, in addition to Heisenberg's position-momentum uncertainty
relations, with quantum excitations.  This idea is illustrated
in the first figure (upper left).  In his 1949 paper, Dirac produced
his light-cone coordinate system as illustrated in the second figure
(upper right).  It is then not difficult to produce the third figure,
for a Lorentz-covariant picture of quantum mechanics. This
Lorentz-squeeze property is observed in high-energy laboratories
through Feynman's parton picture discussed in Sec.~\ref{quarkmo}.
}\label{diracqm33}
\end{figure}

\vspace{1mm}

In each of these papers, Dirac presented the original ingredients
which can serve as building blocks for making quantum mechanics
relativistic.  We combine those elements using Wigner's little
groups~\cite{wig39} and and Feynman's observation of
high-energy physics~\cite{fkr71,fey69a,fey69b}.

\par

First of all, let us combine Dirac's 1945 paper and his light-cone
coordinate system given in his 1949 paper.  Since $x$ and $y$ variables
are not affected by Lorentz boosts along the $z$ direction in
Eq.(\ref{owf02}), it is sufficient to study the Gaussian form
\begin{equation}\label{owf03}
 \exp{\left\{ - \frac{1}{2}\left(z^2 + t^2\right)\right\}} .
\end{equation}
This form is certainly not invariant under Lorentz boost as Dirac noted.
On the other hand, it can be written as
\begin{equation}\label{owf04}
 \exp{\left\{ - \frac{1}{2}\left(u^2 + v^2\right)\right\}} ,\end{equation}
where $u$ and $v$ are the light-cone variables defined in Eq.(\ref{licone01}).
If we make the Lorentz-boost or Lorentz squeeze according to Eq.(\ref{sqz02}),
this Gaussian form becomes
\begin{equation}\label{owf05}
 \exp{\left\{ - \frac{1}{2}\left(e^{-2\eta} u^2 + e^{2\eta} v^2\right)\right\}} .
\end{equation}
If we write the Lorentz boost as
\begin{equation}
z' = \frac{z + \beta t}{\sqrt{1 - \beta^2}}, \qquad
t' = \frac{t + \beta z}{\sqrt{1 - \beta^2}} ,
\end{equation}
where $\beta$ is the the velocity parameter $v/c$, then $\beta$ is related to $\eta$ by
\begin{equation}\label{betaeta}
  \beta = \tanh(\eta).
\end{equation}

\par
Let us go back to the Gaussian form of Eq.(\ref{owf03}), this expression
is consistent with Dirac's earlier paper on the time-energy uncertainty
relation~\cite{dir27}.  According to Dirac, this is a c-number uncertainty
relation without excitations.  The existence of the time-energy uncertainty
is illustrated in the first part of Fig.~\ref{diracqm33}.

\par
In his 1927 paper, Dirac noted the space-time asymmetry in uncertainty
relations.   While there are no time-like excitations, quantum
mechanics allows excitations along the $z$ direction.  How can we
take care of problem?
\par
If we suppress the excitations along the $t$ coordinate, the
normalized solution of this differential equation, Eq.(~\ref{sol02}), is
\begin{equation}\label{sol44}
\psi(z,t) =  \left(\frac{1}{\pi 2^{n}n!} \right)^{1/2}
 H_n(z)\exp{\left\{-\left(\frac{z^2 +t^2}{2}\right)\right\}} .
\end{equation}
If we boost the coordinate system,
the Lorentz-boosted wave functions should take the form
\begin{eqnarray}\label{sol55}
&{}&\psi_{\eta}^n(z,t) = \left(\frac{1}{\pi 2^{n}n!} \right)^{1/2}
 H_n\left(z\cosh\eta - t\sinh\eta\right)     \nonumber\\[2ex]
&{}&  \hspace{5mm} \times
          \exp{\left\{-\left[\frac{(\cosh2\eta)(z^2 + t^2)
               - 4(\sinh2\eta)zt}{2}\right] \right\}}.
\end{eqnarray}
 \par
These are the solutions of the phenomenological equation of Feynman
{\em et al.}~\cite{fkr71} for internal motion of the quarks inside
a hadron.  In 1971, Feynman {\em et al.} wrote
down a Lorentz-invariant differential equation of the form
\begin{equation}\label{diff33}
\frac{1}{2} \left\{-\left(\frac{\partial}{\partial x_{\mu}}\right)^2
 + x_{\mu}^2 \right\} \psi\left(x_{\mu}\right) = (\lambda + 1)
 \psi \left(x_{\mu}\right) ,
\end{equation}
where $x_{\mu}$ is for the Lorentz-covariant space-time four vector.
This oscillator equation is separable in the Cartesian coordinate
system, and the transverse components can be seprated out.  Thus,
the differential of Eq.(\ref{osc05}) contains the essential element
of the Lorentz-invariant Eq.(~\ref{diff33}).
 \par

However, the solutions contained in Ref.~\cite{fkr71} are not normalizable
and therefore cannot carry physical interpretations.  It was shown later
that there are normalizable solutions which constitute a representation
of Wigner's $O(3)$-like little group~\cite{knp86,wig39,kno79}.  The
$O(3)$ group  is the three-dimensional rotation group without a time-like
direction or time-like excitations.  This addresses Dirac's concern about
the space-time asymmetry in uncertainty relations~\cite{dir27}.
Indeed, the expression of Eq.(\ref{sol44}) is considered to be the
representation of Wigner's little group for quantum bound
states~\cite{wig39,kno79}.  We shall return to more physical
questions in Sec.~\ref{quarkmo}.

\section{Further Properties of the Lorentz Harmonics}\label{proba}

Let us continue our discussion of quantum bound states using harmonic
oscillators.  We are interested in this section to see how the oscillator
solution of Eq.(\ref{sol44}) would appear to a moving observer.
\par
The variable $z$ and $t$ are the longitudinal and time-like separations
between the two constituent particles.  In terms
of the light-cone variables defined in Eq.(\ref{licone01}), the solution
of Eq.(\ref{sol44}) takes the form
\begin{equation}\label{cwf11}
 \psi_{0}^{n}(z,t) = \left[\frac{1}{\pi n! 2^{n}} \right]^{1/2}
       H_{n}\left( \frac{u + v}{\sqrt{2}}\right)
       \exp{\left\{-\left(\frac{u^{2} + v^{2}}{2}\right)\right\}} ,
\end{equation}
and
\begin{equation}\label{cwf22}
 \psi_{\eta}^{n}(z,t) = \left[\frac{1}{\pi n! 2^{n}} \right]^{1/2}
       H_{n}\left(\frac{e^{-\eta}u +  e^{\eta} v}{\sqrt{2}}\right)
   \exp{\left\{-\left(\frac{e^{-2\eta}u^{2} + e^{2\eta}v^{2}}{2}
     \right)\right\}} ,
\end{equation}
for the rest and moving hadrons respectively.

\par
It is mathematically possible to expand this as~\cite{knp86,kno79ajp}
\begin{equation}\label{cwf33}
    \psi_{\eta}^{n}(z,t) = \left(\frac{1}{\cosh\eta}\right)^{(n+1)}
     \sum_{k} \left[\frac{(n+k)!}{n!k!}\right]^{1/2}
     (\tanh\eta)^{k}\chi_{n+k}(z)\chi_{n}(t) ,
\end{equation}
where $\chi_n(z)$ is the $n$-th excited state oscillator wave function
which takes the familiar form
\begin{equation}
     \chi_n (z) = \left[\frac{1}{\sqrt{\pi}2^n n!}\right]^{1/2}
              H_n(z) \exp{\left(\frac{-z^2}{2}\right)} ,
\end{equation}
as given in Eq.(\ref{owf01}).
This is an expansion of the Lorentz-boosted wave function in terms of the
Lorentz harmonics.
\par

If the hadron is at rest, there are no time-like oscillations.  There
are time-like oscillations for a moving hadron.  This is the way in
which the space and time variable mix covariantly.  This also provides
a resolution of the space-time asymmetry pointed out by Dirac in his
1927 paper~\cite{dir27}.  We shall return to this question in
Sec.~\ref{restof}.  Our next question is whether those oscillator
equations can be given a probability interpretation.
\par
Even though we suppressed the excitations along the $t$ direction in the
hadronic rest frame, it is an interesting mathematical problem to start
with the oscillator wave function with an excited state in the time
variable.  This problem was adressed by Rotbart in 1981~\cite{rotbart81}.

\subsection{Lorentz-invariant Orthogonality Relations}
Let us consider two wave functions $\psi_{\eta}^n (z,t).$
If two covariant wave functions are in the same Lorentz frame
and have thus the same value of $\eta$, the orthogonality relation
\begin{equation}
\left(\psi^{n'}_{\eta}, \psi^{n}_{\eta}\right) = \delta_{nn'}
\end{equation}
is satisfied.

\par
\begin{figure}
\centerline{\includegraphics[scale=0.56]{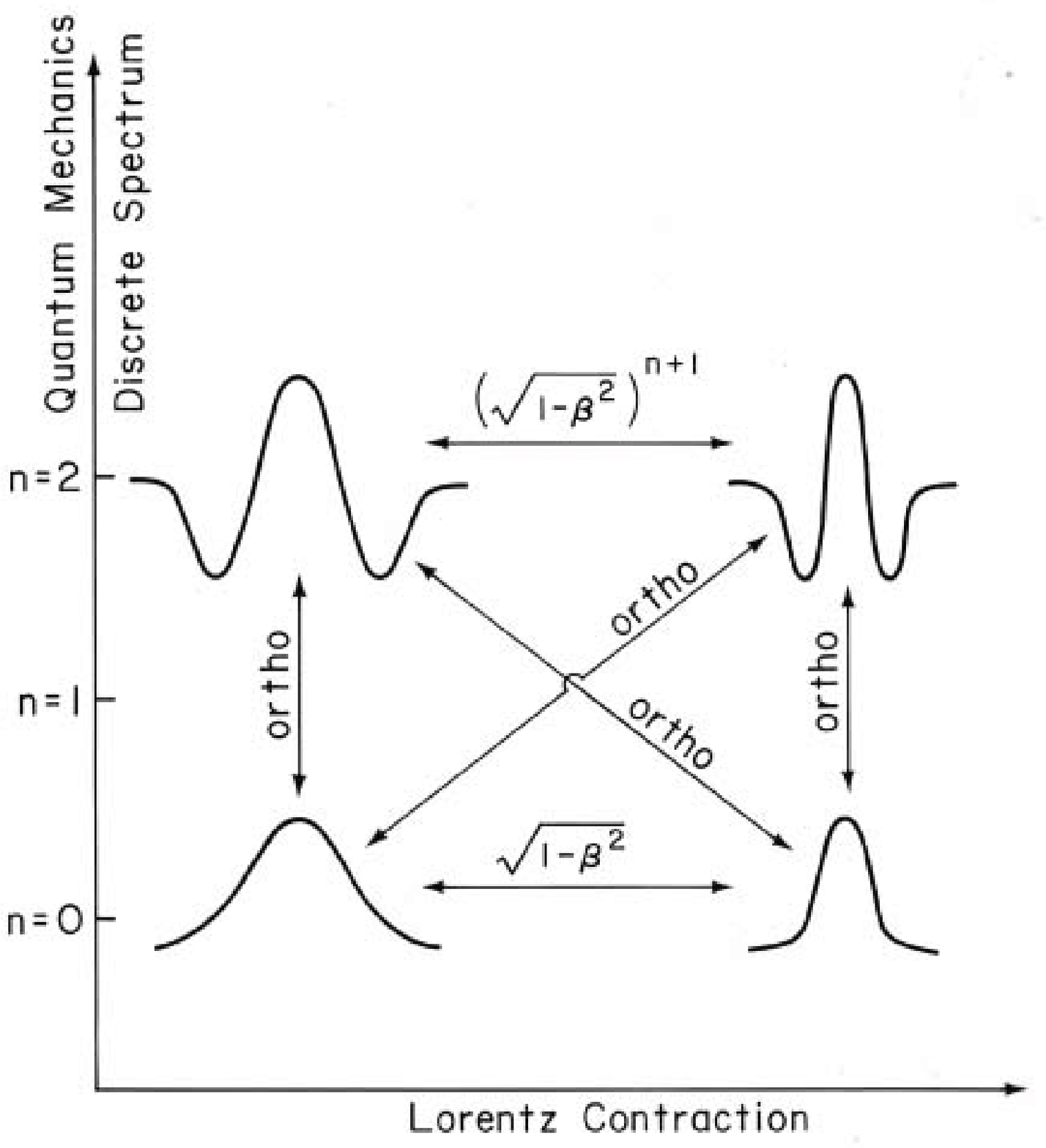}}
\vspace{2mm}
\caption{Orthogonality relations for the covariant harmonic oscillators.
The orthogonality remains invariant.  For the two wave functions in the
orthogonality integral, the result is zero if they have different
values of $n$.  If both wave functions have the same value of $n$, the
integral shows the Lorentz contraction property.
}\label{ortho}
\end{figure}

If those two wave functions have different values of $\eta$, we
have to start with
\begin{equation}
\left(\psi^{n'}_{\eta'}, \psi^{n}_{\eta}\right) =
\int \left(\psi^{n'}_{\eta'}(z,t)\right)^*\psi^{n}_{\eta}(z,t) dz dt .
\end{equation}
Without loss of generality, we can assume $\eta'$ = 0 in the system
where $\eta = 0$, and evaluate the integration.  The result
is~\cite{ruiz74}
\begin{equation}
\left(\psi^{n'}_{0}, \psi^{n}_{\eta}\right) =
\int \left(\psi^{n'}_{0}(z,t)\right)^2 \psi^{n}_{\eta}(z,t) dx dt
         =  \left(\sqrt{1 - \beta^2}\right)^{(n + 1)}\delta_{n,n'} .
\end{equation}
where $\beta = \tanh(\eta),$ as given in Eq.(\ref{betaeta}).
This is like the Lorentz-contraction property of a rigid rod.  The ground
state is like a single rod.  Since we obtain the first excited state
by applying a step-up operator, this state should behave like a
multiplication of two rods, and a similar argument  can be give to $n$
rigid rods. This is illustrated in Fig.~\ref{ortho}.

With these orthogonality properties, it is possible to give quantum
probability interpretation in the Lorentz-covariant world, and it
was so stated in our 1977 paper~\cite{kn77par}.

\subsection{Probability Interpretations}\label{interpre}
Let us study the probability issue in terms of the one-dimensional
oscillator solution of Eq.(\ref{owf01}) whose probability interpretation
is indisputable.   Let us also go back to the rotationally invariant
differential equation of Eq.(\ref{osc03}).  Then the product
\begin{equation}
\chi_{n_x}(x) \chi_{n_y}(y)
\end{equation}
also has a probability interpretation with the eigen value
$\left(n_x + n_y + 1 \right).$   Thus the series of the
form~\cite{knp91,knp86}
\begin{equation}\label{cwf331}
    \phi_{\eta}^{n}(x,y) =
      \left(\frac{1}{\cosh\eta}\right)^{(n+1)}
     \sum_{k} \left[\frac{(n+k)!}{n!k!}\right]^{1/2}
  (\tanh\eta)^{k}\chi_{n+k}(x)\chi_{n}(y)
\end{equation}
also has its probability interpretation, but it is not in an
eigen state.  Each term in this series has an eigenvalue
$(2n + k +1)$.  The expectation value of Eq.(\ref{osc03}) is
\begin{equation}\label{cwf332}
      \left(\frac{1}{\cosh\eta}\right)^{2(n+1)}
     \sum_{k} \frac{(2n + k + 1) (n + k)!}{n!k!}(\tanh\eta)^{2k} .
\end{equation}

If we replace the variables $x$ and $y$ by $z$ and $t$ respectively
in the above expression of Eq.(\ref{cwf331}), it becomes the
Lorentz-covariant wave function of Eq.(\ref{cwf33}).  Each term
$\chi_{n + k} (z) \chi_k (t) $ in the series has the eigenvalue
$n$.  Thus the series is in the eigen state with the eigenvalue $n$.
\par
This difference does not prevent us from importing the probability
interpretation from that of Eq.(\ref{cwf331}).

\par
In the present covariant oscillator formalism, the time-separation
variable can be separated from the rest of the wave function, and does
not requite further interpretation.   For a moving hadron, time-like
excitations are mixed with longitudinal excitations.  Is it possible
to give a physical interpretation to those time-like excitations?
To address this issue, we shall study in Sec.~\ref{squeeze} two-mode
squeezed states also based on the mathematics of Eq.(\ref{cwf331}).
There, both variables have their physical interpretations.

\section{Two-mode Squeezed States}\label{squeeze}

Harmonic oscillators play the central role also in quantum optics.
There the $n^{th}$ excited oscillator state corresponds to the $n$-photon
state $|n>$.  The ground state means the zero-photon or vacuum state
$|0>$.  The single-photon coherent state can be written as
\begin{equation}
 |\alpha> = e^{-\alpha \alpha^{*}/2} \sum_{n} \frac{\alpha^n}{\sqrt{n!}}|n> ,
\end{equation}
which can be written as~\cite{knp91}
\begin{equation}
 |\alpha> = e^{-\alpha \alpha^{*}/2}
    \sum_{n} \frac{\alpha^n}{n!} \left(\hat{a}^{\dagger}\right)^n |0>
  = \left\{e^{-\alpha \alpha^{*}/2}\right\}
     \exp{\left\{\alpha \hat{a}^{\dagger}\right\}} |0> .
\end{equation}
This aspect of the single-photon coherent state is well known.  Here
we are dealing with one kind of photon, namely with a given momentum
and polarization.  The state $|n>$ means there are $n$ photons of
this kind.

\par
Let us next consider a state of two kinds of photons, and write
$|n_1, n_2>$ as the state of $n_1$ photons of the first kind, and
$n_2$ photons of the second kind~\cite{yuen76}.  We can then
consider the form
\begin{equation}\label{coher11}
   \frac{1}{\cosh\eta} \exp{\left\{(\tanh\eta)
    \hat{a}_1^{\dagger}\hat{a}_2^{\dagger} \right\}} |0,0> .
\end{equation}
The operator $\hat{a}_1^{\dagger} \hat{a}_2^{\dagger} $ was studied
by Dirac in connection with his representation of the deSitter group,
as we mentioned in Sec.~\ref{diracqm}.  After making a Taylor
expansion of Eq.(\ref{coher11}), we arrive at
\begin{equation}
    \frac{1}{\cosh\eta} \sum_{k} (\tanh\eta)^k |k, k> ,
\end{equation}
which is the squeezed vacuum state or two-photon coherent
state~\cite{knp91,yuen76}.  This expression is the wave function of
Eq.(\ref{cwf331}) in a different notation.  This form is also called
the entangled Gaussian state of two photons~\cite{gied03} or the
entangled oscillator state of space and time~\cite{kn05job}.

\par
If we start with the $n$-particle state of the first photon, we
obtain
\begin{eqnarray}\label{coher22}
&{}&   \left[\frac{1}{\cosh\eta}\right]^{(n + 1)}
    \exp{\left\{(\tanh\eta) \hat{a}_1^{\dagger}\hat{a}_2^{\dagger}
    \right\}} |n,0>                       \nonumber \\[1ex]
&{}& \hspace{10mm}  = \left[\frac{1}{\cosh\eta}\right]^{(n + 1)}
    \sum_{k} \left[\frac{(n+k)!}{n!k!}\right]^{1/2}
    (\tanh\eta)^k |k + n, k> ,
\end{eqnarray}
which is the wave function of Eq.(\ref{cwf33}) in a different
notation.  This is the $n$-photon squeezed state~\cite{knp91}.

\par
Since the two-mode squeezed state and the covariant harmonic
oscillators share the same set of mathematical formulas, it is
possible to transmit physical interpretations from one to the
other.  For two-mode squeezed state, both photons carry physical
interpretations, while the interpretation is yet to be given to the
time-separation variable in the covariant oscillator formalism.
It is clear from Eq.~(\ref{cwf33}) and Eq.~(\ref{coher22}) that
the time-like excitations are like the second-photon states.
\par
What would happen if the second photon is not observed?  This
interesting problem was addressed by Yurke and Potasek~\cite{yurke87}
and by Ekert and Knight~\cite{ekn89}.   They used the density
matrix formalism and integrated out the second-photon states.
This increases the entropy and temperature of the system.  We
choose not to reproduce their mathematics, because we will
be presenting the same mathematics in Sec.~\ref{restof}.

\section{Time-separation Variable in Feynman's Rest
of the Universe}\label{restof}

As was noted in the previous section, the time-separation variable
has an important role in the covariant formulation of the harmonic
oscillator wave functions.  It should exist wherever the space
separation exists.  The Bohr radius is the measure of the separation
between the proton and electron in the hydrogen atom.  If this atom
moves, the radius picks up the time separation, according to
Einstein~\cite{kn06aip}.

\par
On the other hand, the present form of quantum mechanics does not
include this time-separation variable.  The best way we can
interpret it at  the present time is to treat this time-separation
as a variable in Feynman's rest of the universe~\cite{hkn99ajp}.
In his book on statistical mechanics~\cite{fey72}, Feynman states

\par
\begin{quote}
{\it When we solve a quantum-mechanical problem, what we really do
is divide the universe into two parts - the system in which we are
interested and the rest of the universe.  We then usually act as if
the system in which we are interested comprised the entire universe.
To motivate the use of density matrices, let us see what happens
when we include the part of the universe outside the system.}
\end{quote}

\par

The failure to include what happens outside the system results in
an increase of entropy.  The entropy is a measure of our ignorance
and is computed from the density matrix~\cite{neu32}.  The density
matrix is needed when the experimental procedure does not analyze
all relevant variables to the maximum extent consistent with
quantum mechanics~\cite{fano57}.  If we do not take into account
the time-separation variable, the result is an increase
in entropy~\cite{kiwi90pl,kim07}.

\par
For the covariant oscillator wave functions defined in
Eq.~(\ref{cwf33}), the pure-state density matrix is
\begin{equation}\label{den11}
  \rho_\eta^{n}(z,t;z',t') = \psi_\eta^{n}(z,t) \psi_\eta^{n} (z',t') ,
\end{equation}
which satisfies the condition $\rho^2  = \rho: $
\begin{equation}
  \rho_\eta^{n}(z,t;x',t') = \int \rho_\eta^{n}(z,t;x",t")
   \rho_\eta^{n}(z",t";z',t') dz"dt" .
\end{equation}

\par
However, in the present form of quantum mechanics, it is not possible
to take into account the time separation variables.  Thus, we have to
take the trace of the matrix with respect to the t variable.  Then
the resulting density matrix is
\begin{eqnarray}\label{den22}
&{}& \rho_\eta^{n}(z,z') = \int \psi_\eta^{n}(z,t)
                            \psi_\eta^{n}(z',t) dt \nonumber \\[2ex]
&{}& \hspace{8mm}  = \left(\frac{1}{\cosh\eta}\right)^{2(n + 1)}
     \sum_{k} \frac{(n+k)!}{n!k!}
     (\tanh\eta)^{2k}\psi_{n+k}(z)\psi^*_{n+k}(z') .
\end{eqnarray}
The trace of this density matrix is one, but the trace of $\rho^2$
is less than one, as
\begin{eqnarray}
&{}& Tr\left(\rho^2\right) = \int \rho_\eta^{n}(z,z')
                     \rho_\eta^{n}(z',z) dz dz' \nonumber \\[2ex]
  &{}& \hspace{10mm} = \left(\frac{1}{\cosh\eta}\right)^{4(n + 1)}
  \sum_{k} \left[\frac{(n+k)!}{n!k!}\right]^2 (\tanh\eta)^{4k} ,
\end{eqnarray}
which is less than one.  This is due to the fact that we do not know
how to deal with the time-like separation in the present formulation
of quantum mechanics.  Our knowledge is less than complete.

\par
The standard way to measure this ignorance is to calculate the
entropy defined as
\begin{equation}
          S = - Tr\left(\rho \ln(\rho)\right)  .
\end{equation}
If we pretend to know the distribution along the time-like direction
and use the pure-state density matrix given in Eq.(\ref{den11}),
then the entropy is zero.  However, if we do not know how to deal
with the distribution along $t$, then we should use the density
matrix of Eq.(\ref{den22}) to calculate the entropy, and the result is
\begin{eqnarray}
&{}&   S = 2(n + 1)\left\{(\cosh\eta)^2 \ln(\cosh\eta) -
              (\sinh\eta) \ln(\sinh\eta)\right\} \nonumber \\[1ex]
  &{}& \hspace{10mm} - \left(\frac{1}{\cosh\eta}\right)^{2(n + 1)}
  \sum_{k} \frac{(n+k)!}{n!k!}\ln\left[\frac{(n+k)!}{n!k!}\right]
               (\tanh\eta)^{2k} .
\end{eqnarray}
In terms of the velocity $v$ of the hadron,
\begin{eqnarray}
&{}&  S = -(n + 1)\left\{\ln\left[1 - \left(\frac{v}{c}\right)^2\right]
      + \frac{(v/c)^2 \ln(v/c)^2}{1 - (v/c)^2}   \right\}    \nonumber \\[2ex]
&{}& \hspace{10mm} - \left[1 - \left(\frac{1}{v}\right)^2\right]
  \sum_{k} \frac{(n+k)!}{n!k!}\ln\left[\frac{(n+k)!}{n!k!}\right]
               \left(\frac{v}{c}\right)^{2k} .
\end{eqnarray}

\begin{figure}
\centerline{\includegraphics[scale=0.35]{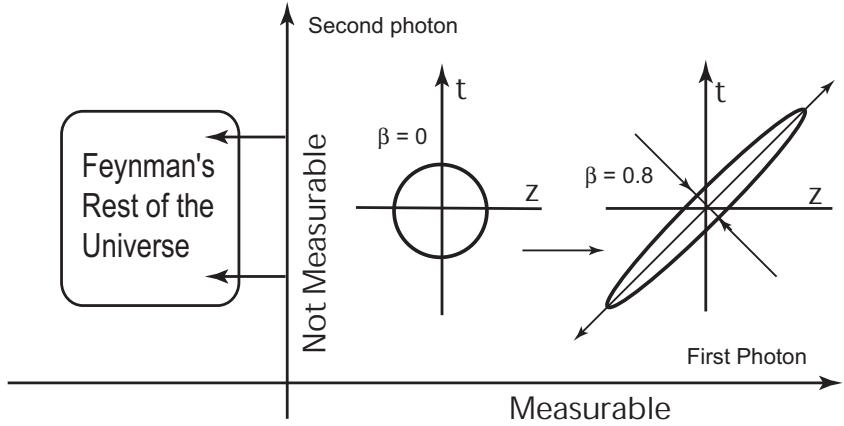}}
\vspace{2mm}
\caption{Localization property in the $zt$ plane.  When the hadron is at
rest, the Gaussian form is concentrated within a circular region specified
by $ (z + t)^2 + (z - t)^2  = 1.$  As the hadron gains speed, the region
becomes deformed to  $ e^{-2\eta}(z + t)^2  + e^{2\eta}(z - t)^2 = 1.$
Since it is not possible to make measurements along the $t$ direction,
we have to deal with information that is less than complete.}\label{restof33}
\end{figure}

\par

Let us go back to the wave function given in Eq.(\ref{cwf22}).
As is illustrated in Fig.~\ref{restof33}, its localization
property is dictated by the Gaussian factor which corresponds to
the ground-state wave function. For this reason, we expect that much
of the behavior of the density matrix or the entropy for the $n^{th}$
excited state will be the same as that for the ground state with
$n = 0.$  For this state, the density matrix and the entropy are
\begin{equation}\label{den33}
  \rho(z,z') = \left(\frac{1}{\pi \cosh(2\eta)}\right)^{1/2}
  \exp{\left\{-\frac{1}{4}\left[\frac{(z + z')^2}{\cosh(2\eta)}
                    + (z - z')^2\cosh(2\eta)\right]\right\}} ,
\end{equation}
and
\begin{equation}
  S = 2\left\{(\cosh\eta)^2 \ln(\cosh\eta) -
             (\sinh\eta)^2 \ln(\sinh\eta)\right\}  ,
\end{equation}
respectively.  The quark distribution $\rho(z,z)$ becomes
\begin{equation}
  \rho(z,z) = \left(\frac{1}{\pi \cosh(2\eta)}\right)^{1/2}
  \exp{\left(\frac{-z^2}{\cosh(2\eta)}\right) }.
\end{equation}

\par
The width of the distribution becomes $\sqrt{\cosh\eta}$, and
becomes wide-spread as the hadronic speed increases.  Likewise,
the momentum distribution becomes wide-spread~\cite{knp86,hkn90pl}.
This simultaneous increase in the momentum and position
distribution widths is called the parton phenomenon in high-energy
physics~\cite{fey69a,fey69b}.  The position-momentum uncertainty
becomes $\cosh\eta$.  This increase in uncertainty is due to our
ignorance about the physical but unmeasurable time-separation
variable.

\par
Let us next examine how this ignorance will lead to the concept
of temperature.  For the Lorentz-boosted ground state with $n = 0$,
the density matrix of Eq.(\ref{den33}) becomes that of the harmonic
oscillator in a thermal equilibrium state if $(\tanh\eta)^2 $ is
identified as the Boltzmann factor~\cite{hkn90pl}.  For other
states, it is very difficult, if not impossible, to describe them
as thermal equilibrium states.  Unlike the case of temperature,
the entropy is clearly defined for all values of $n$.  Indeed,
the entropy in this case is derivable directly from the hadronic
speed.

\par
The time-separation variable exists in the Lorentz-covariant world,
but we pretend not to know about it.  It thus is in Feynman's
rest of the universe.  If we do not measure this time-separation,
it becomes translated into the entropy.

\par
\begin{figure}
\centerline{\includegraphics[scale=0.4]{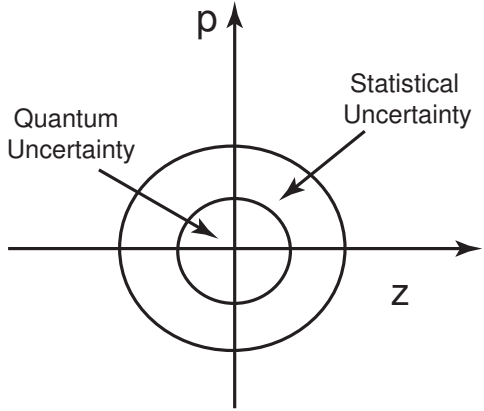}}
\vspace{5mm}
\caption{The uncertainty from the hidden time-separation coordinate.
The small circle indicates the minimal uncertainty when the hadron is
at rest.  More uncertainty is added when the hadron moves.  This is
illustrated by a larger circle.  The radius of this circle increases
by $\sqrt{\cosh(2\eta)}$.}\label{restof44}
\end{figure}

We can see the uncertainty in our measurement process from the
Wigner function defined as
\begin{equation}
W(z,p) = \frac{1}{\pi} \int \rho(z + y, z - y) e^{2ipy} dy .
\end{equation}
After integration, this Wigner function becomes
\begin{equation}
W(z,p) = \frac{1}{\pi\cosh(2\eta)} \exp{\left\{-
        \left(\frac{z^2 + p^2}{\cosh(2\eta)}\right)\right\}} .
\end{equation}
This Wigner phase distribution is illustrated in Fig.~\ref{restof44}.
The smaller inner circle corresponds to the minimal uncertainty
of the single oscillator.  The larger circle is for the total
uncertainty including the statistical uncertainty from our failure
to observe the time-separation variable.  The two-mode squeezed state
tells us how this happens.  In the two-mode case, both the first and
second photons are observable, but we can choose not to observe
the second photon.

\par

\section{Lorentz-covariant Quark Model}\label{quarkmo}

The hydrogen atom played the pivotal role while the present form of
quantum mechanics was developed.  At that time,  the proton was
in the absolute Galilean frame of reference, and it was thinkable
that the proton could move with a speed close to that of light.

\par
Also, at that time, both the proton and electron were point particles.
However, the discovery of Hofstadter {\em et al.} changed the
picture of the proton in 1955~\cite{hofsta55}.  The proton charge has
its internal distribution.  Within the framework of quantum electrodynamics,
it is possible to calculate the Rutherford formula for the electron-proton
scattering when both electron and proton are point particles.
Because the proton is not a point particle, there is a deviation from
the Rutherford formula.  We describe this deviation using the formula
called the ``proton form factor'' which depends on the momentum transfer
during the electron-proton scattering.

\par
Indeed, the study of the proton form factor has been and still is one
of the central issues in high-energy physics.  The form factor decreases
as the momentum transfer increases.  Its behavior is called the
``dipole cut-off'' meaning an inverse-square decrease, and it has been
a challenging problem in quantum field theory and other theoretical
models~\cite{frazer60}. Since the emergence of the quark model in
1964~\cite{gell64}, the hadrons are regarded as quantum bound states
of quarks with space-time wave functions.  Thus, the quark model is
responsible for explaining this form factor.  There are indeed many
papers written on this subject.  We shall return to this problem in
Subsec.~\ref{formfac}.

\par
Another problem in high-energy physics is Feynman's parton
picture~\cite{fey69a,fey69b}.
If the hadron is at rest, we can approach this problem within the
framework of bound-state quantum mechanics.  If it moves with a
speed close to that of light, it appears as a collection of an
infinite number of partons, which interact with external signals
incoherently.  This phenomenon raises the question of whether the
Lorentz boost destroys quantum coherence~\cite{kim98fort}.
This leads to the concept of Feynman's decoherence~\cite{kn03os}.
We shall discuss this problem first.

\begin{figure}
\centerline{\includegraphics[scale=0.5]{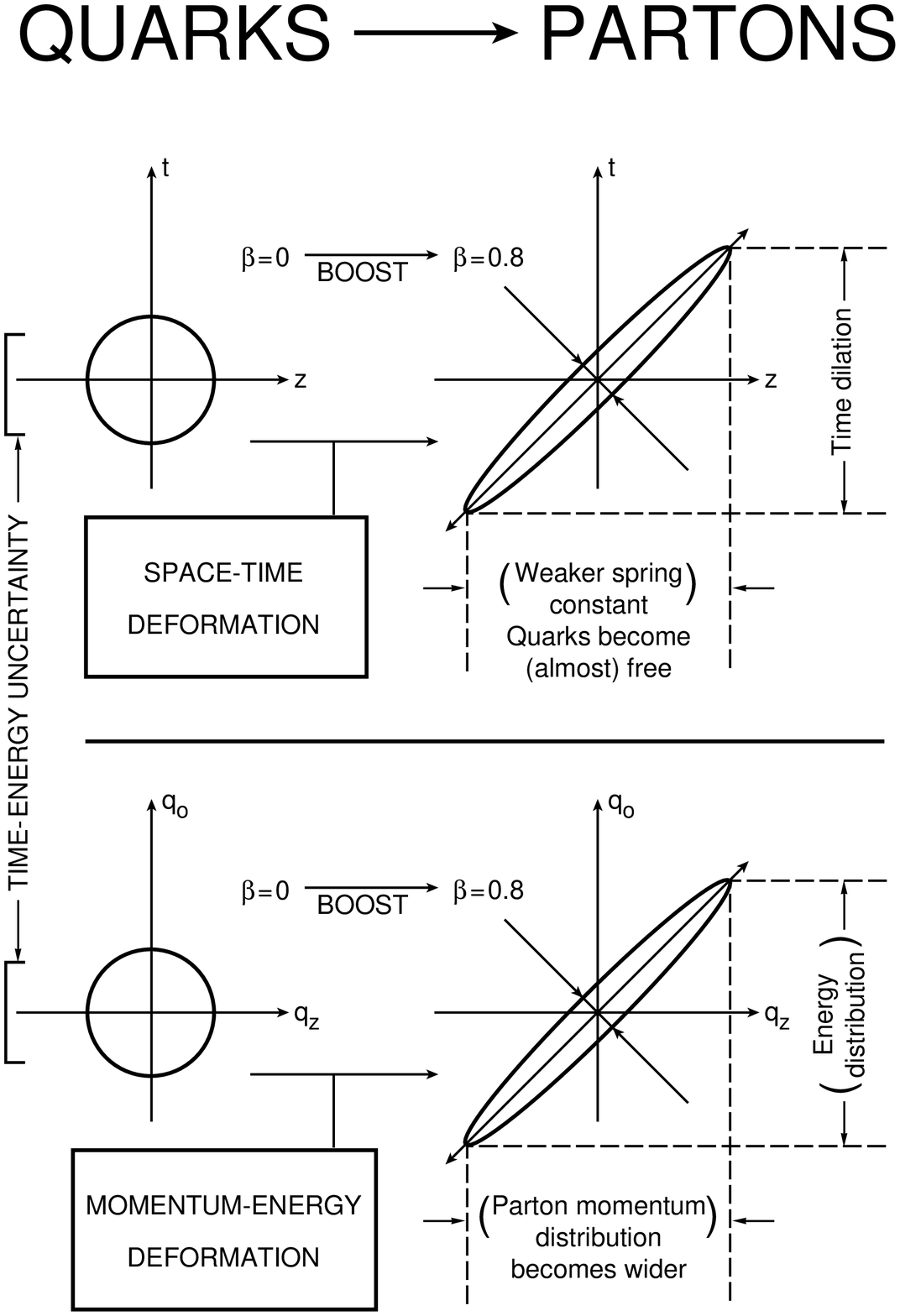}}
\vspace{5mm}
\caption{Lorentz-squeezed space-time and momentum-energy wave
functions.  As the hadron's speed approaches that of light, both
wave functions become concentrated along their respective positive
light-cone axes.  These light-cone concentrations lead to Feynman's
parton picture.}\label{parton}
\end{figure}

\subsection{Feynman's Parton Picture and Feynman's Decoherence}\label{fdeco}

In 1969, Feynman observed that a fast-moving hadron can be regarded
as a collection of many ``partons'' whose properties appear to be
quite different from those of the quarks~\cite{knp86,fey69b}.  For
example, the number of quarks inside a static proton is three, while
the number of partons in a rapidly moving proton appears to be
infinite. The question then is how the proton looking like a bound
state of quarks to one observer can appear different to an observer
in a different Lorentz frame?  Feynman made the following systematic
observations.

\par
\begin{itemize}

\item[a.]  The picture is valid only for hadrons moving with
 velocity close to that of light.

\item[b.]  The interaction time between the quarks becomes dilated,
 and partons behave as free independent particles.

\item[c.]  The momentum distribution of partons becomes widespread as
 the hadron moves fast.

\item[d.]  The number of partons seems to be infinite or much larger
 than that of quarks.

\end{itemize}

\par
\noindent Because the hadron is believed to be a bound state of two
or three quarks, each of the above phenomena appears as a paradox,
particularly b) and c) together.  How can a free particle have a
wide-spread momentum distribution?

\par
In order to address this question, let us go to Fig.~\ref{parton},
which illustrates the Lorentz-squeeze property of the hadron as the
hadron gains its speed. If we use the harmonic oscillator wave
function, its momentum-energy wave function takes the same form as
the space-time wave function.  As the hadron gains its speed, both
wave functions become squeezed.
\par
As the wave function becomes squeezed, the distribution becomes
wide-spread, the spring constant appear to become weaker.
Consequently, the constituent quarks appear to become free
particles.
\par
If the constituent particles are confined in the narrow elliptic
region, they become like massless particles.  If those massless
particles have a wide-spread momentum distribution, it is like
a black-body radiation with infinite number of photon distributions.
\par
We have addressed this question extensively in the literature,
and concluded Gell-Mann's quark model and Feynman's parton model
are two different manifestations of the same Lorentz-covariant
quantity~\cite{kn77par,hussar81,kim89}.  Thus coherent quarks
and incoherent partons are perfectly consistent within the framework
of quantum mechanics and special relativity~\cite{kim98fort}.
Indeed, this defines Feynman's decoherence~\cite{kn03os}.
\par
More recently, we were able to explain this decoherence problem
in terms of the interaction time among the constituent quarks
and the time required for each quark to interact with
external signals~\cite{kn05job}.

\par

\subsection{Proton Form Factors and Lorentz Coherence}\label{formfac}

As early as in 1970, Fujimura {\em et al.} calculated the
electromagnetic form factor
of the proton using the wave functions given in this paper and
obtained the so-called ``dipole'' cut-off of the form
factor~\cite{fuji70}.  At that time, these authors did not
have a benefit of the differential equation of Feynman
and his co-authors~\cite{fkr71}.
Since their wave functions can now be given a bona-fide covariant
probability interpretation, their calculation could be placed
between the two limiting cases of quarks and partons.

\par

Even before the calculation of Fujimura {\it et al.} in 1965,
the covariant wave functions were discussed by various
authors~\cite{yuka53,markov56,ginz65}.  In 1970, Licht and
Pagnamenta also discussed this problem with Lorentz-contracted
wave functions~\cite{licht70}.

\par
In our 1973 paper~\cite{kn73}, we attempted to explain the
covariant oscillator wave function in terms of the coherence
between the incoming signal and the width of the contracted wave
function.  This aspect was explained in terms of the overlap of
the energy-momentum wave function in our book~\cite{knp86}.
\par

In this paper, we would like to go back to the coherence problem
we raised in 1973, and follow-up on it.
In the Lorentz frame where the momentum of the proton has the
opposite signs before and after the collision, the four-momentum
transfer is
\begin{equation}
            (p, E) - (-p, E) = (2p, 0) ,
\end{equation}
where the proton comes along the $z$ direction with its momentum
$p$, and its energy $\sqrt{p^2 + m^2}$.
\par
Then the form factor becomes
\begin{equation}\label{ff11}
F(p) = \int e^{2ipz} \left(\psi_{\eta}(z,t) \right)^* \psi_{-\eta}(z,t)~ dz~ dt .
\end{equation}
If we use the ground-state oscillator wave function, this integral becomes
\begin{equation}
\frac{1}{\pi}
\int e^{2ipz} \exp{\left\{-\cosh(2\eta)\left(z^2 + t^2\right)\right\} }~dz~dt .
\end{equation}
After the $t$ integration, this integral becomes
\begin{equation}
\frac{1}{\sqrt{\pi \cosh(2\eta)}} \int e^{2ipz}
       \exp{\left\{- z^2\cosh(2\eta)\right\} }~dz .
\end{equation}
The integrand is a product of a Gaussian factor and a sinusoidal oscillation.
The width of the Gaussian factor shrinks by $1/\sqrt{\cosh(2\eta)}$, which
becomes $\exp{(-\eta)}$ as $\eta$ becomes large.  The wave length of the
sinusoidal factor is inversely proportional to the momentum $p$.  The wave
length decreases also at the rate of $ \exp{(-\eta}).$  Thus, the rate of
the shrinkage is the same for both the Gaussian and sinusoidal factors.
For this reason, the cutoff rate of the form factor of Eq.(\ref{ff11}) should
be less than that for
\begin{equation}\label{ff22}
\int e^{2ipz} \left(\psi_{0}(z,t) \right)^* \psi_{0}(z,t)~ dz~ dt
 = \frac{1}{\sqrt{\pi}} \int e^{2ipz} \exp{\left(- z^2 \right) }~dz ,
\end{equation}
which corresponds to the form factor without the squeeze effect on
the wave function.  The integration of this expression lead to
$\exp{\left(-p^2\right)},$ which corresponds to an exponential cut-off
as $p^2$ becomes large.

\par
Let us go back to the form factor of Eq.(\ref{ff11}).  If we complete the
integral, it becomes
\begin{equation}
F(p) = \frac{1}{\cosh(2\eta)} \exp{\left\{ \frac{-p^2}
{\cosh(2\eta)} \right\} } .
\end{equation}
As $p^2$ becomes large, the Gaussian factor becomes a constant.  However,
the factor $1/\cosh(2\eta)$ leads the form factor decrease of $1/p^2$,
which is a much slower decrease than the exponential cut-off without
squeeze effect.
\par
There still is a gap between this mathematical formula and the
observed experimental data.  Before looking at the experimental
curve, we have to realize that there are three quarks inside the hadron
with two oscillator mode.  This will lead to a $\left(1/p^2\right)^2$
cut-off, which is commonly called the dipole cut-off in the literature.

\par

There is still more work to be done.  For instance, the effect of
the quark spin should be addressed~\cite{lipes72,henriq75}.  Also
there are reports of deviations from the exact dipole
cut-off~\cite{punjabi05}.  There have been attempts to study the
form factors based on the four-dimensional rotation
group~\cite{roberts05}, and also on the lattice QCD~\cite{matevo05},

\par
\begin{figure}
\centerline{\includegraphics[scale=0.5]{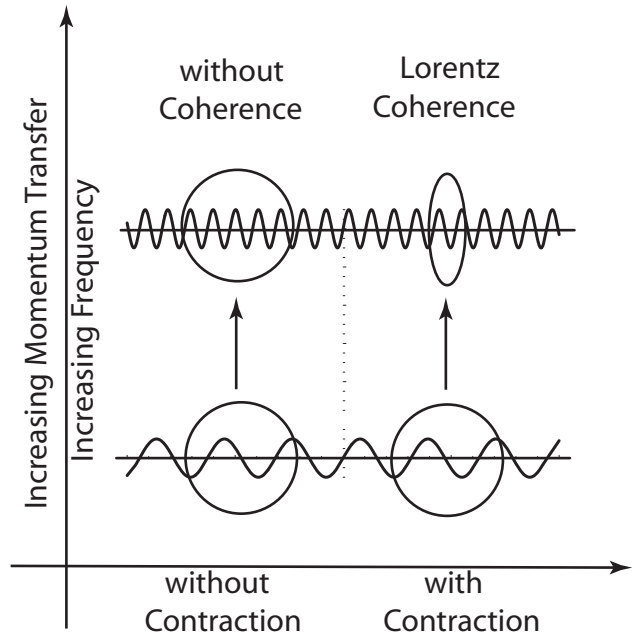}}
\vspace{5mm}
\caption{Coherence between the wavelength and the proton size.  As the
momentum transfer increases, the external signal sees Lorentz-contracting
proton distribution.  On the other hand, the wavelength of the signal
also decreases.  Thus, the cutoff is not as severe as the case where
the proton distribution is not contracted.
}\label{locoher}
\end{figure}

Yet, it is gratifying to note that the effect of Lorentz squeeze
lead to the polynomial decrease in the momentum transfer,
thanks to the Lorentz coherence illustrated in Fig.~\ref{locoher}.
We started our logic from the fundamental principles of quantum
mechanics and relativity.

\section*{Conclusions}

In this paper, we presented one mathematical formalism applicable both
to the entanglement problems in quantum optics~\cite{gied03} and to
high-energy hadronic physics~\cite{kn05job}.  The formalism is based
on harmonic oscillators familiar to us.  We have presented a complete
orthonormal set with a Lorentz-covariant probability interpretation.
\par
Since both branches of physics share the same mathematical base,
it is possible to translate physics from one branch to the other.
In this paper, we have given a physical interpretation to the
time-separation variable as a hidden variable in Feynman's rest
of the universe, in terms of the two-mode squeezed state where
both photons are observable.
\par
This paper is largely a review paper with an organization to suit
the current interest in physics.  For instance, the concepts of
entanglement and decoherecne did not exist when those original
papers were written.  Furthermore, the probability interpretation
given in Subsection~\ref{interpre} has not been published before.
\par
The rotation symmetry plays its role in all branches of physics.
We noted that the squeeze symmetry plays active roles in two
different subjects of physics.  It is possible that the squeeze
transformation can serve useful purposes in many other fields,
although we are not able to specify them at this time.

\end{document}